\documentclass[pre,aps,preprint,showpacs,preprintnumbers,amsmath,amssymb,secnumroman,eqsecnum]{revtex4}

\usepackage{graphicx}
\usepackage{dcolumn}
\usepackage{bm}

\newcommand{\beq}{\begin{equation}}
\newcommand{\eeq}{\end{equation}}
\newcommand{\beqa}{\begin{eqnarray}}
\newcommand{\eeqa}{\end{eqnarray}}
\newcommand{\vc}[1]{\mbox{\boldmath $#1$}}

\newcommand{\vol}[1]{{\bf #1}}

\newcommand{\du}[1]{{\bf\sf #1}}

\begin{document}


\title{Swimming at small Reynolds number of a collinear assembly of spheres in an incompressible viscous fluid with inertia}

\author{B. U. Felderhof}

 \email{ufelder@physik.rwth-aachen.de}
\affiliation{Institut f\"ur Theorie der Statistischen Physik \\ RWTH Aachen University\\
Templergraben 55\\52056 Aachen\\ Germany\\
}%

\date{\today}

\begin{abstract}
Swimming at small Reynolds number of a collinear assembly of identical spheres immersed in an incompressible viscous fluid is studied on the basis of a set of equations of motion for the individual spheres. The motion of the spheres is caused by actuating forces and forces derived from a direct interaction potential, as well as hydrodynamic forces exerted by the fluid  as frictional and added mass hydrodynamic interactions. The swimming velocity is deduced from the momentum balance equation for the assembly of spheres, and the mean power required during a period is calculated from an instantaneous power equation. Expressions are derived for the mean swimming velocity and the mean power, valid to second order in the amplitude of displacements from the relative equilibrium positions. Hence these quantities can be evaluated in terms of prescribed periodic displacements. Explicit calculations are performed for a linear chain of three identical spheres.

\end{abstract}

\maketitle
\section{\label{I}Introduction}

The general theory of swimming and flying must incorporate the effects of both friction and inertia \cite{1}. Swimming in the full range of scale number, varying between the Stokes limit, dominated by viscosity, and the inertial limit, dominated by the mass of the spheres and the mass density of the fluid, was studied elsewhere for a deformable sheet \cite{2}, a slab \cite{2A}, and a sphere \cite{3}. The inertial limit has been discussed in general terms for a deformable body of arbitrary shape \cite{4}. Models involving recoil locomotion due to shifting internal mass \cite{5},\cite{6} provide other examples of swimming by inertial effect.

In earlier work we investigated the effect of fluid inertia on the motion of a collinear swimmer consisting of a chain of rigid spheres immersed in a viscous incompressible fluid \cite{7},\cite{8}. The analysis was based on a set of approximate equations of motion for the individual spheres incorporating direct interaction forces, as well as frictional and added mass hydrodynamic interactions. The model allows study of the whole range of scale number. Vortex shedding \cite{9},\cite{10} is neglected in the model.

In the following we study small amplitude motion of the model system in more detail. In Sec. II we specify the basic model equations and derive the balance equation for the total momentum of the spheres, as influenced by interaction with the fluid, as well as an instantaneous power equation \cite{11}. In Sec. III we specialize to small amplitude motion and derive the corresponding equations of the bilinear theory. As application we consider in Sec. IV the case of a linear chain of three identical neutrally buoyant spheres interacting via harmonic springs.

In the bilinear theory of swimming the mean swimming velocity is expressed as the expectation value of a speed matrix with respect to the vector of displacements of the sphere centers from their equilibrium positions. Similarly, the mean power is expressed as the expectation value of a power matrix, derived from the mean rate of dissipation. In earlier work \cite{7},\cite{8} we used a matrix representation based on the transformation from Cartesian to center and relative coordinates. We show here that it is advantageous to use instead a representation based on the eigenvectors of the elastic matrix. This incorporates the symmetries of the system and for the three-sphere swimmer leads to relatively simple expressions, which are presented in analytic form.

The mean swimming speed can be optimized for given power by a suitable choice of stroke. The optimization leads to a generalized eigenvalue problem in terms of the speed matrix and the power matrix. It turns out that for the three-sphere swimmer the maximum eigenvalue and the corresponding eigenvector hardly depend on the scale number. As a consequence the optimal motion found in the Stokes limit is nearly optimal in the whole range of scale number, including the inertial limit.

\section{\label{II}Dynamics of collinear spheres}

We consider a system of $N$ identical spheres of radius $a$ and mass density $\rho_p$ immersed in a viscous incompressible fluid of shear viscosity $\eta$ and mass density $\rho$. The fluid is of infinite extent in all directions. We assume that at all times the centers of the spheres are collinear and located on the $x$ axis of a Cartesian system of coordinates. The dynamics of the system is governed by an interaction potential $V_{int}$, depending on the instantaneous configuration of centers, by actuating forces $(\vc{E}_1(t),...,\vc{E}_N(t))$, directed in the $x$ direction and summing to zero total force at any time $t$, and by hydrodynamic interactions transmitted by the fluid. We assume that the hydrodynamic interactions can be approximated by Stokes friction, calculated from the Stokes equations, and by added mass effects, calculated from potential theory. We summarize the positions of centers in the $N$-dimensional vector $\du{R}=(x_1,...,x_N)$, and the sphere momenta in $\du{p}=(p_1,...,p_N)$. The momenta are related to the velocities $\du{U}=(U_1,...,U_N)$ by
\begin{equation}
\label{2.1}\du{p}=\du{m}\cdot\du{U},
\end{equation}
where $\du{m}$ is the mass matrix, which depends on the relative positions of the sphere centers, so that it is invariant under translations of the whole assembly. The dynamics of the system is assumed to be governed by the approximate equations of motion \cite{7}
\begin{equation}
\label{2.2}\frac{d\du{R}}{dt}=\du{U},\qquad\frac{d\du{p}}{dt}=-\frac{\partial\mathcal{K}}{\partial\du{R}}-\vc{\zeta}\cdot\du{U}-\frac{\partial V_{int}}{\partial\du{R}}+\du{E},
\end{equation}
where the kinetic energy $\mathcal{K}$ is given by
\begin{equation}
\label{2.3}\mathcal{K}=\frac{1}{2}\;\du{p}\cdot\du{w}\cdot\du{p},
\end{equation}
with inverse mass matrix $\du{w}=\du{m}^{-1}$. The derivative with respect to positions in Eq. (2.2) is to be taken at constant momenta $\du{p}$. The friction matrix $\vc{\zeta}$ and the interaction potential $V_{int}$ are invariant under translations of the assembly. We abbreviated $\du{E}=(E_1,...,E_N)$.

In the absence of actuating forces the system comes to rest due to friction with the fluid. The rest situation corresponds to a solution of Eq. (2.2) with constant configuration $\du{R}_0$, which is a minimum of the potential energy $V_{int}$. In the rest configuration the center and relative positions are
\begin{equation}
\label{2.4}C_0=\frac{1}{N}\sum^N_{j=1}x_{0j},\qquad c_{0j}=x_{0j}-C_0,\qquad j=1,...,N.
\end{equation}
In shorthand notation $\du{c}_0=\du{R}_0-\du{C}_0$. From the definitions it follows that
\begin{equation}
\label{2.5}\du{u}\cdot\du{c}_0=0,
\end{equation}
where $\du{u}=(1,1,...,1)$.

The sphere positions are summarized in the $N$-vector
\begin{equation}
\label{2.6}\du{R}(t)=\du{C}_0+\du{c}_0+\du{u}\int^t_0U(t')\;dt'+\du{d}(t),
\end{equation}
where $U(t)$ is the swimming velocity and $\du{d}(t)$ are the additional displacements from relative equilibrium positions. By definition $\du{u}\cdot\du{d}=0$.
The sphere velocities are summarized in the $N$-vector
\begin{equation}
\label{2.7}\du{U}=U\du{u}+\dot{\du{d}}.
\end{equation}
Substituting this into Eq. (2.2) and requiring that the total actuating force vanishes we obtain an equation of motion for $U(t)$, involving also the time-derivatives $\dot{\du{d}}$ and $\ddot{\du{d}}$ of the displacements $\du{d}(t)$.

The requirement that the sum of actuating forces vanishes reads in abbreviated notation
\begin{equation}
\label{2.8}\du{u}\cdot\du{E}=0.
\end{equation}
We can use this requirement to derive a simple balance equation for the total momentum
 \begin{equation}
\label{2.9}P=\du{u}\cdot\du{p}.
\end{equation}
The total momentum varies in time due to interaction with the fluid. From Eq. (2.2) we derive
 \begin{equation}
\label{2.10}\frac{dP}{dt}=-\du{u}\cdot\vc{\zeta}\cdot\du{U},
\end{equation}
The kinetic energy and potential energy terms in Eq. (2.2) do not contribute on account of translational invariance.
The actuating forces $\du{E}(t)$ do not occur explicitly in Eq. (2.10) and this allows a kinematic point of view in which the velocity $U(t)$ is determined from the equation for prescribed displacements $\du{d}(t)$.

We define the Hamiltonian $\mathcal{H}$ as
 \begin{equation}
\label{2.11}\mathcal{H}=\mathcal{K}+V_{int}.
\end{equation}
From Eq. (2.2) we find for its time-derivative
 \begin{equation}
\label{2.12}\frac{d\mathcal{H}}{dt}=-\mathcal{D}+\du{E}\cdot\du{U},
\end{equation}
with rate of dissipation
 \begin{equation}
\label{2.13}\mathcal{D}=\du{U}\cdot\vc{\zeta}\cdot\du{U}.
\end{equation}
This may be called the instantaneous power equation \cite{11}.

In periodic swimming the time-average of the rate of dissipation over a period equals the power used. We denote the average as
 \begin{equation}
\label{2.14}\overline{\mathcal{D}}=\frac{1}{\tau}\int^\tau_0\mathcal{D}(t)\;dt,
\end{equation}
where $\tau$ is the period. From Eq. (2.12) we see that the mean rate of dissipation equals the power, i.e. the work performed by the actuating forces during a period,
 \begin{equation}
\label{2.15}\overline{\mathcal{D}}=\overline{\du{E}\cdot\du{U}}.
\end{equation}
In the same way we see from Eq. (2.10)
 \begin{eqnarray}
\label{2.16}\du{u}\cdot\overline{\vc{\zeta}\cdot\du{U}}=0.
\end{eqnarray}
This shows that in periodic swimming the mean drag vanishes.

\section{\label{III}Bilinear theory}
In the following we consider a collinear assembly with small deviations from an equilibrium configuration $\du{c}_0$. The averaged Eq. (2.16) is solved by formal expansion in powers of the displacements $\du{d}(t)$. We include terms up to second order. The displacements are assumed to vary harmonically in time at frequency $\omega=2\pi/\tau$.

To second order Eq. (2.16) reads
\begin{equation}
\label{3.1}\du{u}\cdot\overline{\vc{\zeta}^{(1)}\cdot\du{U}^{(1)}}+\du{u}\cdot\vc{\zeta}^0\cdot\overline{\du{U}^{(2)}}=0,
\end{equation}
where $\vc{\zeta}^0$ is the friction matrix of the equilibrium configuration, which is time-independent.
From Eq. (3.1) we find for the mean second order swimming velocity
\begin{equation}
\label{3.2}Z^0\overline{U^{(2)}}=\overline{\mathcal{I}_{T}^{(2)}},
\end{equation}
with friction coefficient $Z^0=\du{u}\cdot\vc{\zeta}^0\cdot\du{u}$ and mean second order translational impetus
\begin{equation}
\label{3.3}\overline{\mathcal{I}^{(2)}_{T}}=-\du{u}\cdot\overline{\vc{\zeta}^{(1)}\cdot\du{U}^{(1)}}.
\end{equation}

The first order friction matrix can be expressed as
\begin{equation}
\label{3.4}\vc{\zeta}^{(1)}=\du{d}\cdot\nabla\vc{\zeta}\big|_0,
\end{equation}
where $\nabla$ is the gradient operator in $N$-dimensional configuration space. From Eqs. (2.6) and (2.7) we find for the first order velocity vector and the corresponding position vector
\begin{equation}
\label{3.5}\du{R}^{(1)}=C^{(1)}\du{u}+\du{d},\qquad\du{U}^{(1)}=U^{(1)}\du{u}+\dot{\du{d}}.
\end{equation}
Here we use a first order equation derived from Eq. (2.10),
\begin{equation}
\label{3.6}\mathcal{M}^0\frac{dU^{(1)}}{dt}+\du{u}\cdot\du{m}^0\cdot\ddot{\du{d}}=-Z^0U^{(1)}-\du{u}\cdot\vc{\zeta}^0\cdot\dot{\du{d}},
\end{equation}
with mass coefficient $\mathcal{M}^0=\du{u}\cdot\du{m}^0\cdot\du{u}$.
This may be regarded as a linear response equation determining the first order velocity $U^{(1)}$ in terms of the displacements $\du{d}$. The equation can be solved by Fourier analysis. The equation for the complex Fourier coefficients reads
\begin{equation}
\label{3.7}\big[-i\omega\mathcal{M}^0+Z^0\big]U^{(1)}_{\omega}=\du{u}\cdot[\omega^2\du{m}^0+i\omega\vc{\zeta}^0]\cdot\du{d}_\omega.
\end{equation}
The solution of this equation can be expressed as
\begin{equation}
\label{3.8}U^{(1)}_{\omega}=i\omega Y(\omega)\du{f}(\omega)\cdot\du{d}_\omega=i\omega\vc{\Psi}(\omega)\cdot\du{d}_\omega,
\end{equation}
with admittance $Y(\omega)=(-i\omega\mathcal{M}^0+Z^0)^{-1}$ and impedance vector
\begin{equation}
\label{3.9}\du{f}(\omega)=(-i\omega\du{m}^0+\vc{\zeta}^0)\cdot\du{u}.
\end{equation}
The elements of the $N$-vector $\vc{\Psi}(\omega)=Y(\omega)\du{f}(\omega)$ are dimensionless.

In the calculation of the time-averaged impetus in Eq. (3.3) we encounter bilinear expressions. The average is evaluated conveniently in complex notation. For example
 \begin{equation}
\label{3.10}\overline{\du{d}U^{(1)}}=\frac{1}{2}\;\mathrm{Re}\;i\omega\du{d}^*_\omega\vc{\Psi}(\omega)\cdot\du{d}_\omega.
\end{equation}
The leading contribution to the mean second order impetus in Eq. (3.3) takes the form
\begin{equation}
\label{3.11}-\du{u}\cdot\overline{(\du{d}\cdot\nabla\vc{\zeta})\cdot\dot{\du{d}}}=\frac{1}{2}\;\mathrm{Re}\;[i\omega\du{d}^*_\omega\cdot\du{D}\big|_0\cdot\du{d}_\omega],
\end{equation}
with derivative friction matrix
\begin{equation}
\label{3.12}\du{D}=\vc{\nabla}\du{f},\qquad\du{f}=\vc{\zeta}\cdot\du{u}=\du{u}\cdot\vc{\zeta},
\end{equation}
as introduced earlier \cite{12}. We write the complete expression as
\begin{equation}
\label{3.13}\overline{\mathcal{I}_{T}^{(2)}}=\frac{1}{2}\;\mathrm{Re}\;[i\omega\du{d}^*_\omega\cdot\breve{\du{D}}\big|_{\du{R}_0}\cdot\du{d}_\omega],
\end{equation}
where the $N\times N$ matrix $\breve{\du{D}}$ differs from $\du{D}$ by a correction coming from $U^{(1)}_\omega$ in Eq. (3.5),
\begin{equation}
\label{3.14}\breve{\du{D}}(\omega)=\du{D}-\du{g}\vc{\Psi}(\omega),\qquad\du{g}=\du{D}\cdot\du{u}.
\end{equation}

The time-dependent rate of dissipation is given by Eq. (2.13). The second order mean rate of dissipation can be expressed in terms of periodic displacements with Fourier amplitude $\du{d}_\omega$ as
\begin{equation}
\label{3.15}\overline{\mathcal{D}^{(2)}}=\frac{1}{2}\;\omega^2\mathrm{Re}\;[\du{d}_\omega^*\cdot\breve{\vc{\zeta}}(\omega)\big|_{\du{R}_0}\cdot\du{d}_\omega],
\end{equation}
with modified friction matrix \cite{7},\cite{13}
\begin{equation}
\label{3.16}\breve{\vc{\zeta}}(\omega)=\vc{\zeta}-\du{f}(\omega)\vc{\Psi}(\omega).
\end{equation}

Finally we note that in Eqs. (3.13) and (3.15) the matrices $i\omega\breve{\du{D}}(\omega)$ and $i\omega\breve{\vc{\zeta}}(\omega)$ can be reduced to their hermitian part. Also we can use a projector to take account of the fact that the displacement vector $\du{d}_\omega$ must be orthogonal to $\du{u}$.

\section{\label{IV}Three-sphere chain}

As an application of the equations derived in the preceding sections we consider swimming of a linear chain of three spheres of radius $a$ and mass density $\rho$. For simplicity we consider neutrally buoyant spheres. As basic equilibrium configuration $\du{c}_0$ we consider
\begin{equation}
\label{4.1}\du{c}_0=(-d,0,d).
\end{equation}

We take the potential energy to be given by the expression
\begin{equation}
\label{4.2}V_{int}(\du{R})=\frac{1}{2}\;k\big[(r_{12}-d)^2+(r_{23}-d)^2\big],
\end{equation}
with elastic constant $k$ and relative distances
\begin{equation}
\label{4.3}r_{12}=x_2-x_1,\qquad r_{23}=x_3-x_2.
\end{equation}
The potential energy is positive definite, translation-invariant, and it vanishes at configuration $\du{c}_0$.

We assume that the mobility matrix $\vc{\mu}(\du{R})$ is given by Oseen hydrodynamic interactions \cite{14} and that the inverse mass matrix $\du{w}(\du{R})$ is evaluated in dipole approximation \cite{7}. The kinetic energy $\mathcal{K}(\du{R},\du{p})$ is positive definite and translation-invariant. The friction matrix and the mass matrix depend only on the relative distances given by Eq. (4.3). The friction coefficient and the total mass in the equilibrium configuration take the values
 \begin{eqnarray}
\label{4.4}Z^0&=&18\pi\eta ad\frac{4d-7a}{4d^2+3ad-18a^2},\nonumber\\
\mathcal{M}^0&=&2\pi\rho a^3d^3\frac{24d^3-31a^3}{8d^6+a^3d^3-16a^6}.
\end{eqnarray}

We introduce the projection operators $\du{P}_{op}$ and $\du{Q}$ defined as
\begin{equation}
\label{4.5}\du{P}_{op}=\frac{1}{3}\;\du{u}\du{u},\qquad \du{Q}=\du{I}-\du{P}_{op}.
\end{equation}
The displacement vector $\du{d}_\omega$ must satisfy
\begin{equation}
\label{4.6}\du{d}_\omega=\du{Q}\cdot\du{d}_\omega,
\end{equation}
to exclude rigid body translation. The projected vector has only two independent components. This allows reduction of the matrices $\breve{\du{D}}(\omega)$ and $\breve{\vc{\zeta}}(\omega)$ in Eqs. (3.14) and (3.16) to a two-dimensional representation.

In order to construct the reduced matrices we expand the displacement vector $\du{d}_\omega$ in terms of a convenient set of basis vectors. We use the orthonormal set of eigenvectors of the elasticity matrix corresponding to the interaction potential $V_{int}$. The latter can be expressed as
\begin{equation}
\label{4.7}V_{int}=\frac{1}{2}\;(\du{R}-\du{R}_0)\cdot\du{H}\cdot(\du{R}-\du{R}_0),
\end{equation}
with $3\times 3$ elasticity matrix $\du{H}$ given explicitly by
 \begin{equation}
\label{4.8}\du{H}=k\left(\begin{array}{ccc}1&-1&0\\-1&2&-1\\0&-1&1
\end{array}\right).
\end{equation}
This has the orthonormal set of eigenvectors
 \begin{eqnarray}
\label{4.9}\du{e}_1&=&(1,1,1)/\sqrt{3}=\du{u}/\sqrt{3},\nonumber\\
\du{e}_2&=&(-1,0,1)/\sqrt{2},\nonumber\\
\du{e}_3&=&(1,-2,1)/\sqrt{6},
\end{eqnarray}
with eigenvalues
\begin{equation}
\label{4.10}\lambda_1=0,\qquad\lambda_2=k,\qquad\lambda_3=3k.
\end{equation}
The first eigenvector and eigenvalue correspond to free translation.

The basis vectors $\du{e}_2,\du{e}_3$ span the reduced vector space. The corresponding displacement vector
\begin{equation}
\label{4.11}\du{d}_\omega=b_2\du{e}_2+b_3\du{e}_3
\end{equation}
is characterized by two complex coefficients $b_2,b_3$ in complex notation.

The matrix $\du{D}$ does not depend on frequency. We can express the matrix $\du{D}$ at $\du{R}_0$ as
 \begin{equation}
\label{4.12}\du{D}=d_{21}\du{e}_2\du{e}_1+d_{23}\du{e}_2\du{e}_3+d_{32}\du{e}_3\du{e}_2,
\end{equation}
with coefficients
 \begin{eqnarray}
\label{4.13}d_{21}&=&6\sqrt{6}\pi\eta a^2\frac{20d^2-72ad+63a^2}{(4d^2+3ad-18a^2)^2},\nonumber\\
d_{23}&=&-12\sqrt{3}\pi\eta a^2\frac{2d^2+9a^2}{(4d^2+3ad-18a^2)^2},\nonumber\\
d_{32}&=&36\sqrt{3}\pi\eta a^2\frac{4d-9a}{16d^3-81a^2d+54a^3}.
\end{eqnarray}
Hence the reduced $2\times 2$ matrix $\du{D}_2$
is given by
 \begin{equation}
\label{4.14}\du{D}_2=\left(\begin{array}{cc}0&d_{23}\\d_{32}&0
\end{array}\right).
\end{equation}
The vectors $\du{g}$ and $\vc{\Psi}(\omega)$ in Eq. (3.14) are given by
 \begin{equation}
\label{4.15}\du{g}=g_2\du{e}_2,\qquad\vc{\Psi}(\omega)=\frac{1}{\sqrt{3}}\;\du{e}_1+\sqrt{\frac{2}{3}}\;S\;\du{e}_3,
\end{equation}
with coefficients
 \begin{eqnarray}
\label{4.16}g_2&=&\sqrt{3}\;d_{21},\nonumber\\ S&=&a\;\frac{9(8d^6+a^3d^3-16a^6)-14is^2a^2d^2(4d^2+3ad-18a^2)}{9(4d-7a)(8d^6+a^3d^3-16a^6)-2is^2d^2(4d^2+3ad-18a^2)(24d^3-31a^3)}.\nonumber\\
\end{eqnarray}
Hence the reduced $2\times 2$ matrix $\breve{\du{D}}_2$
is given by
 \begin{equation}
\label{4.17}\breve{\du{D}}_2=\left(\begin{array}{cc}0&d_{23}-\sqrt{2}d_{21}S\\d_{32}&0
\end{array}\right).
\end{equation}
From Eq. (3.13) we define correspondingly
\begin{equation}
\label{4.18}\du{B}_2(\omega)=\frac{ia}{2Z^0}\;\big(\breve{\du{D}}_2-\breve{\du{D}}_2\;^\dagger\big).
\end{equation}
The matrix $\du{B}_2(\omega)$ is hermitian and its elements are dimensionless.

We define the hermitian matrix $\du{A}(\omega)$ as
\begin{equation}
\label{4.19}\du{A}(\omega)=\frac{1}{2\eta a}\big[\breve{\vc{\zeta}}(\omega)+\breve{\vc{\zeta}}(\omega)^\dagger\big].
\end{equation}
The elements of $\du{A}(\omega)$ are dimensionless. For the three-sphere swimmer in the representation given by Eq. (4.9) the elements of the first row and column of the three-dimensionless matrix $\du{A}(\omega)$ vanish. Therefore we can restrict attention to the $2\times 2$ matrix $\du{A}_2(\omega)$ obtained by deleting the first row and column. The reduced matrix $\du{A}_2(\omega)$ is diagonal of the form
\begin{equation}
\label{4.20}\du{A}_2(\omega)=\left(\begin{array}{cc}A_{222}&0\\0&A_{233}\end{array}\right),
\end{equation}
with elements
\begin{eqnarray}
\label{4.21}A_{222}&=&\frac{24\pi d}{4d-3a},\nonumber\\
A_{233}&=&72\pi d\frac{27(4d-7a)(8d^6+a^3d^3-16a^6)^2+4s^4d^4(4d^2+3ad-18a^2)W}{81(4d-7a)^2(8d^6+a^3d^3-16 a^6)^2+4s^4d^4(4d^2+3ad-18a^2)^2(24d^2-31a^2)^3},\nonumber\\
\end{eqnarray}
with coefficient
\begin{equation}
\label{4.22}W=192d^7+480ad^6-496a^3d^4-1296a^4d^3+353a^6d+816a^7.
\end{equation}
The element $A_{222}$ is remarkably simple and does not depend on $s$. The element $A_{233}$ depends in complicated fashion on the ratio $d/a$, but its dependence on the scale number $s$ is quite simple.

The optimal stroke for swimming in the $x$ direction for given power is given by the eigenvector of the generalized eigenvalue problem
\begin{equation}
\label{4.23}\du{B}_2(\omega)\cdot\du{b}=\lambda\du{A}_2(\omega)\cdot\du{b},
\end{equation}
with maximum eigenvalue $\lambda_{max}$.  The maximum eigenvalue equals the efficiency $E_T=\eta\omega a^2|\overline{U^{(2)}}|/\overline{\mathcal{D}^{(2)}}$ of the stroke $\du{b}$. It has the dimensionless value
\begin{equation}
\label{4.24}\lambda_{max}=\frac{a}{2Z^0}\;\frac{|d_{23}-d_{32}-\sqrt{2}\;d_{21}S|}{\sqrt{A_{222}A_{233}}}.
\end{equation}
As a function of $s$ this takes the form
\begin{equation}
\label{4.25}\lambda_{max}(s)=\sqrt{\frac{A+Bs^4}{1+Cs^4}},
\end{equation}
with real coefficients $A,B,C$ which depend only on the ratio $d/a$. In the Stokes limit $s=0$ the value $\lambda_{max}(0)$ agrees with the analytic expression $\lambda_+$ derived earlier \cite{12}. In Fig. 1 we plot the ratio $\lambda_{max}(s)/ \lambda_{max}(0)$ as a function of $s$ for $d=3a$. To lowest order in the ratio $a/d$
\begin{equation}
\label{4.26}\lambda_{max}(s)=\frac{7}{48\pi\sqrt{3}}\;\frac{a^2}{d^2}+O\big(\frac{a^3}{d^3}\big),
\end{equation}
independent of $s$.

The eigenvector $\du{b}$ corresponding to the maximum eigenvalue takes the form
\begin{equation}
\label{4.27}\du{b}=(1,b_3),\qquad b_3=\frac{-2iZ^0}{a}\frac{A_{222}}{d_{23}-d_{32}-\sqrt{2}d_{21}S}\;\lambda_{max},
\end{equation}
when normalized such that the first component equals unity. In Fig. 1 we plot the real and imaginary parts of the second component $b_3$ for $d=3a$ as functions of $s$. Both are nearly constant, in agreement with an earlier plot \cite{7}. To lowest order in $a/d$ the second component $b_3=i$, independent of $s$.

The time-dependent displacement for the stroke $\varepsilon\du{b}$ is given by
\begin{equation}
\label{4.28}\du{d}(t)=\mathrm{Re}\;[\varepsilon(\du{e}_2+b_3\du{e}_3)\exp(-i\omega t)].
\end{equation}
As shown above, for the optimal stroke this hardly depends on the scale number $s$, so that over the whole range of $s$ the motion is very similar to that in the Stokes limit. Nonetheless for large $s$ the motion of the center of mass is dominated by inertial effects, as is evident from Eqs. (3.8) and (3.9) of Ref. 8.

\section{\label{IX}Discussion}

The bilinear theory of swimming is important because it allows an optimization procedure for finding the small amplitude stroke of maximum mean swimming velocity for given power. We showed elsewhere \cite{7},\cite{8} for the example of a chain of three identical spheres that for larger amplitude of the same stroke the efficiency is nearly the same. In the above we have presented the equations of the bilinear theory for the three-sphere chain in analytic form. In particular this allows a detailed analysis of the dependence of mean swimming speed and power on the scale number. In the model the full range of viscosity from the Stokes limit, relevant to microorganisms, to the inertial limit of small viscosity and reactive swimming, is covered, and the transition between the two types of swimming can be studied. We found that the optimal stroke and efficiency hardly vary over the full range of scale number.

For the collinear chain the analysis of the swimming velocity can be based on the simple momentum balance equation Eq. (2.10). In the equation the actuating forces do not occur, and this allows a kinematic point of view. One can expect that for more complicated geometry a similar analysis based on equations for total momentum and angular momentum is feasible. We have implemented such an analysis for planar assemblies of spheres. The study of fully three-dimensional structures of spheres remains an open problem.

In the model the phenomenon of vortex shedding \cite{10},\cite{11} is absent, and it would be desirable to extend the model to take account of this effect. A first goal would be to explain the interesting experimental and computer simulation findings of Klotsa et al. \cite{15}  for a collinear two-sphere system.

\newpage

\section*{Figure captions}

\subsection*{Fig. 1}
Plot of the ratio $\lambda_{max}(s)/\lambda_{max}(0)$ (drawn curve), as well as $\mathrm{Re}\;[b_3]$ (short dashes) and $\mathrm{Im}\;[b_3]$ (long dashes) as functions of scale number $s$ for $d=3a$.

\newpage
\setlength{\unitlength}{1cm}
\begin{figure}
 \includegraphics{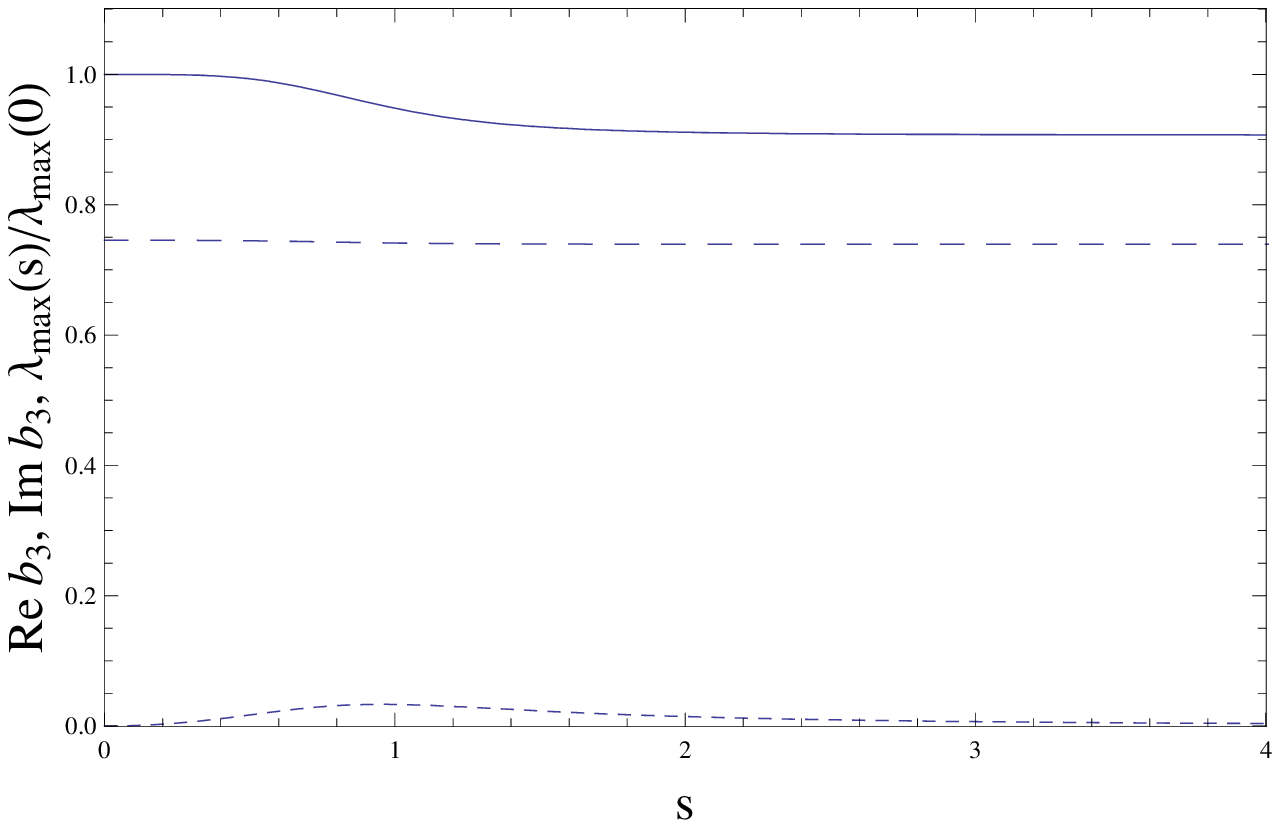}
   \put(-9.1,3.1){}
\put(-1.2,-.2){}
  \caption{}
\end{figure}
\newpage


\newpage

\begin{thebibliography}{99}

\bibitem{1}
S. Childress, {\it Mechanics of swimming and flying"}, (Cambridge University Press, Cambridge, 1981).

\bibitem{2}
E. O. Tuck,  "A note on a swimming problem", J. Fluid Mech. \vol{31}, 305 (1968).

\bibitem{2A}
B. U. Felderhof,  "Swimming of a deformable slab in a viscous incompressible fluid with inertia", arXiv:1507.01186[physics.flu-dyn].

\bibitem{3}
B. U. Felderhof and R. B. Jones, "Swimming of a sphere in a viscous incompressible fluid with inertia", arXiv:1512.04667[physics.flu-dyn].

\bibitem{4}
R. Mason and J. Burdick, "Propulsion and control of deformable bodies in an ideal fluid",
 in {\it Robotics and Automation} , Proceedings IEEE Int. Conf., \vol{1}, 773 (1999).

\bibitem{5}
P. G. Saffman, "The self-propulsion of a deformable body in a perfect fluid", J. Fluid Mech. \vol{28}, 385 (1967).

\bibitem{6}
S. Childress, S. E. Spagnolie, and T. Tokieda, "A bug on a raft: recoil locomotionin a viscous fluid", J. Fluid Mech. \vol{669}, 527 (2011).

\bibitem{7}
B. U. Felderhof, "Effect of inertia on laminar swimming and flying of an assembly of rigid spheres in an incompressible viscous fluid", Phys. Rev. E \vol{92}, 053011 (2015).

\bibitem{8}
B. U. Felderhof, "Effect of fluid inertia on the motion of a collinear swimmer", arXiv:1606.02400[physics.flu-dyn].

\bibitem{9}
W. Shyy, Y. Lian, J. Tang, D. Viieru, and H. Liu, {\it Aerodynamics of low Reynolds number flyers}, Cambridge University Press, Cambridge, 2011).

\bibitem{10}
J. Peng and J. O. Dabiri, "An overview of a Lagrangian method for analysis of animal wake dynamics", J. Exp. Biol. \vol{211}, 280 (2008).

\bibitem{11}
G. J. van Ingen Schenau and P. R. Cavanagh, "Power equations in endurance sports", J. Biomech. \vol{23}, 865 (1990).

\bibitem{12}
B. U. Felderhof, "Swimming of an assembly of rigid spheres at low Reynolds number", Eur. Phys. J. E \vol{37}, 110 (2014).

\bibitem{13}
B. U. Felderhof, "Swimming of a linear chain with a cargo in an incompressible viscous fluid with inertia", arXiv:1607.08048[physics.flu-dyn].

\bibitem{14}
H. Yamakawa, {\it Modern Theory of Polymer Solutions} (Harper and Row, New York, 1971).

\bibitem{15}
D. Klotsa, K. A. Baldwin, R. J. A. Hill, R. M. Bowley, and M. R. Swift, "Propulsion of a two-sphere swimmer", Phys. Rev. Lett. \vol{115}, 248102 (2015).

\end{thebibliography}
\end{document}